\documentclass[reprint,aps,prl,twocolumn,
]{revtex4-2}
\usepackage{amsmath}
\usepackage{amsfonts}
\usepackage{amssymb}
\usepackage{mathrsfs}
\usepackage{bbm}
\usepackage{graphicx}
\usepackage{dcolumn}
\usepackage{bm}
\usepackage{hyperref}% add hypertext capabilities
\usepackage{pstricks}
\usepackage{pst-3dplot}
\usepackage{subfigure}
\usepackage{xy}
\usepackage{comment}
\usepackage{enumerate}
\usepackage{xcolor}

\usepackage[normalem]{ulem} %\sout{}

\definecolor{SU}{rgb}{.3,0,.3}

\definecolor{dg}{rgb}{.01,.6,.006} 
\newcommand{\changed}[1]{#1}

\newcommand{\sectionPRL}[1]{\paragraph{#1---\!\!\!\!\!}}
\newcommand{\subsectionPRL}[1]{\paragraph{#1---\!\!\!\!\!}}

\begin{document}

\title{Entanglement of quantum systems via a 
classical mediator\\ in hybrid van Hove theory}

\author{Sebastian Ulbricht$^{1,2}$}
\email{sebastian.ulbricht@ptb.de}
\author{Andr\'{e}s Dar\'{i}o Berm\'{u}dez Manjarres$^{3}$}
\author{Marcel Reginatto$^{1}$}

\affiliation{$^1$Physikalisch-Technische Bundesanstalt, Bundesallee 100, 38116 Braunschweig, Germany\\
$^2$Institut für Mathematische Physik, Technische Universität Braunschweig,
 Mendelssohnstraße 3, 38106 Braunschweig, Germany\\
$^3$Universidad Distrital Francisco Jos\'{e} de Caldas, Cra. 7 No. 40B-53, Bogot\'{a}, Colombia}

\date{\today}

\begin{abstract}
It is a matter of ongoing discussion whether quantum states can become entangled while only interacting via a classical mediator. This lively debate is deeply interwoven with the question of whether entanglement studies can prove the quantum nature of gravity. However, the answer to this fundamental question depends crucially on which hybrid quantum-classical theory is used. In this Letter, we demonstrate that entanglement by a classical mediator is possible within the framework of hybrid van Hove theory, showing that existing no-go theorems on that matter do not universally apply to hybrid theories in general. After briefly recapitulating the key features of the hybrid van Hove theory, we show this using the example of two quantum spins coupled by a classical harmonic oscillator. By deriving the spin density matrix for this scenario and comparing it to its equivalent for a pure quantum system, we show that entanglement between the two spins is generated in both cases. Conclusively, this is illustrated by presenting the purity and concurrence of the spin-spin system as a decisive measure for entanglement. Our results further imply that quantum entanglement studies cannot rule out consistent quantum theories featuring classical gravity.   
\end{abstract}

\maketitle

\sectionPRL{Introduction} Entanglement, as first emphasized by Schr\"{o}dinger \cite{Schroedinger_1935}, is one of the characteristic features of quantum mechanics \cite{4xHorodecki2009}. In the context of hybrid theories, where quantum and classical systems interact, it is natural to ask whether two quantum systems that do not interact directly can become entangled via a classical mediator. We cannot answer this question without making assumptions about the nature of classical and quantum systems, how they are modeled, and, equally important, the way they interact. Thus, as one would expect, the answer necessarily depends on the details of the hybrid theory used to model the interacting classical-quantum system \cite{HallReginatto2018}. This is not just an academic question, as various no-go theorems 
\cite{Bose2017,MarlettoVedral2017,Galley2022}
claim that the answer is negative and relevant to experimental tests of the quantum nature of gravity.
This claim, however, has \changed{been disputed and} not gone unchallenged \cite{HallReginatto2018,
Anastopoulos2022,Doner2022,Pal2021}\changed{\cite{Aziz2025,DiBiagio2025}}. \changed{It is the primary intend of this Letter to contribute to this ongoing discussion from the perspective of the new approach of} Hybrid van Hove (HvH) mechanics \cite{Bermudez2024,Reginatto2025}.
In addition, this theory is also relevant to computations of a complex quantum system where part of it is approximated using classical physics; in this case, it is essential to determine whether such an approximation is adequate to capture correlations between quantum subsystems that may interact indirectly.
 
\begin{figure}[t!]
    \vspace{-2em}\centering
    \includegraphics[scale = 0.15]{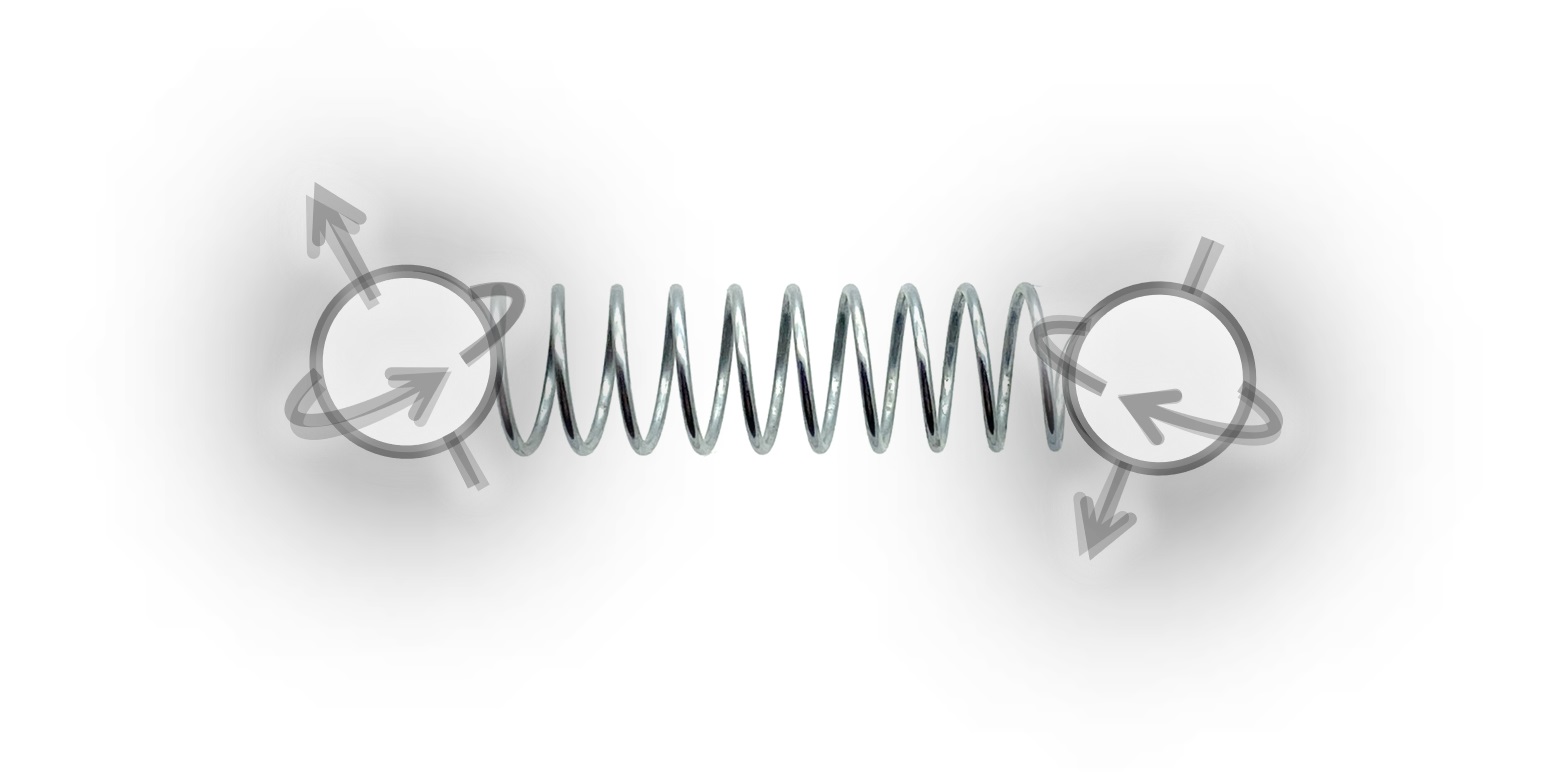}\vspace{-3em}\\
    \caption{In HvH theory, two quantum spins can become entangled while only interacting indirectly via a classical oscillator as the mediator.}
\label{fig:front_picture}\vspace{-2em}
\end{figure}

In this paper, we consider entanglement via a 
classical mediator in the framework of Hybrid van Hove mechanics and carry out detailed calculations for a relatively simple system of two spins, or qubits, that interact via a classical oscillator, see Fig.~\ref{fig:front_picture}. The theory is formulated in Hilbert space, and, as we explicitly show, it allows for entanglement via a classical mediator. This result contradicts various no-go theorems in the literature, indicating that some of the assumptions of these theorems are too restrictive and, in some cases, may be even inconsistent with any formulation of classical mechanics in Hilbert space that incorporates one of its most defining features, namely the Poisson algebra of phase space functions.

A hybrid theory of interacting classical and quantum systems requires a common mathematical framework that can include both types of systems. HvH mechanics is formulated in Hilbert space, with Schr\"{o}dinger operators representing quantum observables, whereas classical observables are represented by the operators introduced by van Hove \cite{VanHove1951}, leading to a formalism that is fundamentally different from the better-known approach of Koopman-von Neumann (KvN) \cite{K1931,vN1930} and its extension to hybrid systems, pioneered by Sudarshan \cite{Sudarshan1976,Sudarshan1978,Sudarshan1979a,Sudarshan1979b}.

\sectionPRL{Hybrid van Hove theory}
 %we proposed a new hybrid theory based on van Hove operators and classical physics described in Hilbert space 
%\block{(a) \emph{classicality:}}
The main difference between our hybrid theory and the usual formulations of hybrid systems in Hilbert space lies in the way the classical sector is described. 
 We introduce a notion of \textit{classicality} with the following requirements for the wavefunction $\phi(\mathbf{q},\mathbf{p},t)$ of the classical state and the operators $\hat{{\mathcal O}}_F$ that represent classical observables of phase space functions $F(\mathbf{q},\mathbf{p})$:\\ 
1. The commutator algebra of observables is isomorphic to the Poisson algebra of phase space functions,
        \begin{equation}\label{Iso_vH_CM}  [\hat{\mathcal{O}}_{F},\hat{\mathcal{O}}_{G}]=i\hbar\hat{\mathcal{O}}_{\{F,G\}},
        \end{equation}
        where ${\{\cdot,\cdot\}}$ is the usual Poisson bracket \footnote{The appearance of $\hbar$ in the Eqs. (\ref{Iso_vH_CM}) and  (\ref{vH_operator}) is not linked to any quantization procedure: the set of unitary representations of the group of contact transformations of van Hove constitutes a continuous family of representations labeled by a real parameter that van Hove calls $\alpha$ and we have chosen $\alpha=1/\hbar$ because it has special properties \cite{VanHove1951}.}.\\
2. The density $\varrho:=|\phi|^2$ satisfies the Liouville equation.\\
3. The expectation value of $\hat{\mathcal{O}}_{F}$ satisfies
        \begin{equation}\label{ExpValOperator} \langle\phi|\hat{\mathcal{O}}_{F}|\phi\rangle = \int dq dp \, \varrho(q,p) F(q,p).
        \end{equation}
It is straightforward to check that the  first requirement is satisfied by the operators
\begin{equation}
\hat{{\mathcal O}}_F =  F-p\,\partial_p F + i\hbar\left( \partial_q F \,\partial_p - \partial_p F \,\partial_q  \right) \label{vH_operator}\,,
\end{equation}
\changed{here in their representation as phase space differential operators, as} introduced by van Hove \cite{VanHove1951}.
Deriving the Schr\"{o}dinger-like equation $i\hbar\frac{\partial\phi}{\partial t}=\hat{{\mathcal O}}_H\phi$ for the van Hove operator of the Hamilton function $H(q,p)=\frac{p^2}{2m}+V(q)$ acting on $\phi=\sqrt{\varrho}e^{i\sigma/\hbar}$, we find
\begin{eqnarray}  
\frac{\partial\mathcal{\varrho}}{\partial t}+\{\varrho,H\} &=& \quad\frac{d\varrho}{dt} \quad=\quad 0, \label{LE}\\
\varrho \left(  \frac{\partial\sigma}{\partial t}+\{\sigma,H\} \right) &=& \varrho \, \frac{d\sigma}{d t} ~\quad=\quad \varrho L\label{dsigmadt}\,.
\end{eqnarray}
Thus, $\varrho=|\phi|^2$ obeys the Liouville equation (\ref{LE}) and therefore fulfills the second requirement. Moreover, Eq. (\ref{dsigmadt}) implies that $\sigma$ is the \textit{classical action}, given by the time integral of the Lagrange function $L(q,p)=\frac{p^2}{2m}-V(q)$.

As we discuss in Ref.~\cite{Bermudez2024}, the third requirement imposes a constraint on the classical wave function $\phi$, namely that the conditions
\begin{equation}\label{constr}
\nabla_q \sigma \simeq\mathbf{p},\quad \nabla_p \sigma\simeq0,\quad \frac{\partial \sigma}{\partial t}\simeq-H, 
\end{equation}
must be identically satisfied (``$\simeq$") when the derivatives of $\sigma$ are evaluated \textit{along any classical trajectory}.
These conditions can always be implemented by setting
\begin{equation}\label{sigmaConstr}
	\sigma =  \eta + H[\tau- \tau'-t]\,, 
\end{equation}	
where $\eta(q,p)$ and $\tau(q,p)$ satisfy the relations $\{\eta,H\}=L$ and $\{\tau,H\}=1$, respectively. The latter relation implies $d\tau/dt$=1, which integrates to $\tau\simeq \tau'+t$ when evaluated over the trajectory starting at the point $q',p'$, provided we set $\tau'=\tau(q',p')$. See the End Matter for more technical details about the interdependency of the phase (\ref{sigmaConstr}) and the conditions (\ref{constr}).

It must be noted that there is a crucial difference between the van Hove operators $\hat{{\mathcal O}}_F$ from (\ref{vH_operator}), characterizing a classical system, and the well-known operators $\hat F$ of quantum mechanics. 
Here we will only briefly recapitulate them. For a more detailed discussion, see Ref.~\cite{Reginatto2025}.

On the one hand, the commutator algebra $[\hat{F},\hat{G}]$ of operators in quantum mechanics is not isomorphic to the Poisson algebra of phase space functions $\{F,G\}$, a well-known consequence of the Groenewold-van Hove theorem \cite{G1999}.
In contrast, this isomorphism is a requirement, satisfied by van Hove operators, as seen in Eq.~(\ref{Iso_vH_CM}).
This difference provides us with a clear, algebraic distinction between classical and quantum observables, as these two sets of operators satisfy inequivalent Lie algebras. \changed{We want to stress again, that by this fact the classicality of the system is preserved and no `effective' quantization or pre-quantization procedure is applied by the reformulation of classical physics in Hilbert space using the van Hove operators}.

On the other hand, the operators of quantum mechanics constitute a product algebra and, in particular, allow for powers of operators $(\hat F)^n$ representing powers of phase space functions $F^n$. This is not the case for van Hove operators.
 Using Eq.~(\ref{vH_operator}), one can check that $\hat{\mathcal{O}}_{F^n}\not=(\hat{\mathcal{O}}_F)^n$ by direct calculation. As an important consequence, \textit{there is no uncertainty principle for classical states despite the non-commutativity of $\hat{{\mathcal{O}}}_q$ and $\hat{{\mathcal O}}_p$} \cite{Bermudez2024,Reginatto2025}.

Having discussed van Hove operators and the way in which they differ from operators in quantum mechanics, we can combine both to construct a mixed quantum-classical theory. In the HvH theory, we consider interacting classical-quantum systems defined via a hybrid Hamiltonian operator
\begin{equation}
\hat{\mathcal{H}} = {\hat{\mathcal{O}}}_{H_C} + {\hat{H}}_Q + \hat{\mathcal{W}}, \label{HybHamil}
\end{equation}
where the van Hove operator $\hat{\mathcal{O}}_{H_C}$ is the Hamiltonian of the classical sector and $\hat{H}_Q$ is the quantum Hamiltonian operator. The coupling term allows for classical-quantum interactions, for instance of the form $\hat{\mathcal{W}}={\hat{\mathcal{O}}}_A\hat{B}$ for some van Hove operator ${\hat{\mathcal{O}}}_A$ and a quantum operator $\hat{B}$.

HvH systems satisfy a number of consistency conditions \cite{Reginatto2025}. For example, one can show that $[\hat{\mathcal{O}}_{F},\hat{G}]=0$, implying that a measurement of a classical observable cannot detect whether or not a local transformation has been applied to the quantum sector, and vice versa. Thus, non-local signaling is excluded. It is also straightforward to establish conservation laws: energy is conserved, as are all observables represented by operators that commute with the hybrid Hamiltonian operator.

Finally, we emphasize that many no-go theorems about hybrid systems do not apply to HvH theory; namely, those that assume that classical observables must be represented by commuting operators and those that assume that they must form a product algebra, as these assumptions do not hold for van Hove operators.

\sectionPRL{Two spins coupled by a harmonic oscillator} 
Here, we study two spins, i.e. qubits, interacting via a harmonic oscillator.
We first consider spin-spin interactions in the framework of standard quantum mechanics. Afterward, we compare the results with those obtained from a quantum-classical hybrid system in which the harmonic oscillator is considered an inherently classical object.

\subsectionPRL{Quantum case}
In a standard representation of a system of two spins, we assume the spin basis \mbox{$|\uparrow\rangle=(1,0)$} and \mbox{$|\downarrow\rangle=(0,1)$} for each spin and construct the product state basis, using the tensor product (i.e $|\uparrow\rangle\otimes|\uparrow\rangle =(1,0,0,0)=:|\uparrow\uparrow\rangle$, and so on). 
%\block{(b) \emph{Hamiltonian}:} 
We further introduce the Hamiltonian operator 
\begin{eqnarray}
    \hat H &=& H_0(\hat q,\hat p)+\frac{1}{2}\left(\epsilon+g\hat q\right)\left(
    \sigma_3\otimes\mathbb{I}+ \mathbb{I}\otimes\sigma_3
    \right) \label{eqn:Hamiltonian}
\end{eqnarray}		
of the spins, coupled to a quantized harmonic oscillator, emerging from the Hamilton function $H_0(q,p)=\frac{p^2}{2m}+\frac{1}{2}m\omega^2  q^2$ by $(p,q)\to(\hat p,\hat q)$.
The interaction term in the Hamiltonian (\ref{eqn:Hamiltonian}) is constructed from standard Pauli spin matrices, coupled to the position of the mediating particle. It is widely studied, also in the context of hybrid theories \cite{Pedernales2022,GonzalezConde2023}.
In the spin product state basis we have $\left(
    \sigma_3\otimes\mathbb{I}+ \mathbb{I}\otimes\sigma_3
    \right)/2=\mathrm{diag}(1,0,0,-1)$ and get four Schrödinger equations
\begin{eqnarray}
 i\hbar\dot \psi_k &=& \left(-\frac{\hbar^2}{2m}\partial^2_q+\frac{1}{2}m\omega^2 q^2+\epsilon_k+g_kq\right) \psi_k \label{eqn:q_Schroedinger_equ2}
\end{eqnarray}
  for the four spin-basis functions $\psi=(\psi_1,\psi_2,\psi_3,\psi_4)$ in the position space representation $(\hat p,\hat q)=(-i\hbar\partial_q,q)$, only differing in 
 $(\epsilon_k)=(\epsilon,0,0,-\epsilon)$, i.e,  $(g_k)=(g,0,0,-g)$.
By completing the square \footnote{Completing the square 
$\frac{1}{2}m\omega^2 q^2+g_kq
 = \frac{1}{2}m\omega^2 \left(q+\frac{g_k}{m \omega^2}\right)^2-\frac{g_k^2}{2m\omega^2}$
gives rise to a shift of energy and position in the Schrödinger equation (\ref{eqn:q_Schroedinger_equ2})} 
 in Eq.~(\ref{eqn:q_Schroedinger_equ2}), for each $k$ we obtain the Schrödinger equation $i\hbar \dot\phi_k=H_0(\hat q,\hat p) \phi_k$ for the standard harmonic oscillator 
 wave function $\phi_k(q,t)$, but displaced in position space and with an offset energy:
\begin{equation}
 \psi_k(q,t)=\mathrm{e}^{-\frac{i}{\hbar}\left(\epsilon_k-\frac{g_k^2}{2m\omega^2}\right)t}\, \phi_k\left(q+\frac{g_k}{m \omega^2},t\right)\,.
\end{equation}
For this example calculation we want the oscillator to be initially prepared in a coherent state $\phi_c(q,t;q_0,p_0)$ with initial position $q_0$ and momentum $p_0$, regardless of the state of the spins. The well-known explicit expression for a coherent oscillator quantum state is recalled in the End Matter of this Letter.
This can be interpreted as the scenario where spins and oscillator are independent at $t=0$ and start to interact for $t>0$, or just as a measurement of the oscillator's position and momentum as the starting point of the scenario (destroying all entanglement that may have existed before). 
To satisfy this requirement, the initial condition has to be shifted separately for each of the spin wave function components: 
\begin{equation}   \phi_k\left(q+\frac{g_k}{m \omega^2},t\right)=\phi_c\left(q+\frac{g_k}{m \omega^2},t;q_0+\frac{g_k}{m \omega^2},p_0\right).
\end{equation}
The spin density matrix is obtained by marginalization of the quantum harmonic oscillator. Its components are given by the integrals
\begin{equation}  \rho_{ij}(t)=\int\mathrm{d}q\,\psi^\ast_i(q,t)\psi_j(q,t)\,.\label{eqn:q_margnial}
\end{equation}
The Gaussian integral (\ref{eqn:q_margnial}) can be solved straightforwardly using standard methods, resulting in a reduced density matrix for the spin states that can be represented in the 
\emph{Bloch-Fano decomposition} \cite{Bengtsson2006,Friis2017}
\begin{eqnarray}
    \rho=\frac{1}{4}\left(\mathbb{I}_4+A^i\cdot(\sigma_i\otimes\mathbb{I}+\mathbb{I}\otimes\sigma_i)+T^{ij}\sigma_i\otimes\sigma_j\right) \label{eqn:dm_01}
\end{eqnarray}
with $i=1,2$ (no $\sigma_3$ involved),  the (identical) Bloch vectors $\vec{A}=\left(\mathrm{Re}(b),\mathrm{Im}(b)\right)$, and the correlation tensor 
\begin{eqnarray}
T=\frac{1}{2}\left(\begin{array}{cc}1+\mathrm{Re}(a) & \mathrm{Im}(a)\\\mathrm{Im}(a) &1-\mathrm{Re}(a)\end{array}\right)\,, \label{eqn:dm_03}
\end{eqnarray}
where the parameters $a=\mathrm{e}^{-2R-iU}$ and $b=\mathrm{e}^{-R/2-iU/2-iS}$ are determined by the functions
\begin{eqnarray}
 R^{(q)}(t)&=&\frac{g^2}{m \omega ^3 \hbar }(1-\cos ( \omega t)) \label{eqn:qdm_Function_01}\\
 S^{(q)}(t)&=&\frac{1}{2}\frac{g ^2 }{ m \omega ^3 \hbar }(\sin (\omega t )-\omega t)\\
 U^{(q)}(t)&=&\frac{g}{m \omega ^2 \hbar}\!\{2 m \omega x_0   \sin (\omega t) \label{eqn:qdm_Function_03}\\
  & &  \quad\quad\quad+p_0 [1\!-\!2 \cos (\omega t)]\}+2\epsilon t/\hbar\nonumber\,,
\end{eqnarray}
depending on the interaction parameters and on the properties of the oscillator. $R$ determines the absolute value of $a$ and $b$, while the other two determine their phases. Notice that the initial conditions $x_0$ and $y_0$ only enter via the function $U$.

Next, we will calculate the mixed quantum classical system in the hybrid van Hove theory.
We will see that this leads to the same form (\ref{eqn:dm_01}) - (\ref{eqn:dm_03}) of the density matrix, however, with slightly different functions $R$, $S$, and $U$. 
Therefore, we introduced the superscript `$(q)$' in Eqs.~(\ref{eqn:qdm_Function_01}) -- (\ref{eqn:qdm_Function_03}), to later compare them with their hybrid counterparts. 

\subsectionPRL{Hybrid case}
We follow the same steps as in the previous section, however, now considering the oscillator to be classical. Within the hybrid van Hove theory, the Hamiltonian $H\to \hat {\mathcal{O}}_H$ is represented by the van Hove-Operator according to Eq.~(\ref{vH_operator}).
A Schrödinger-type equation governs the dynamics of the resulting mixed system 
\begin{eqnarray}
& &i\hbar\frac{\partial\Psi^{(h)}_{k}}{\partial t}=   \left[-\frac{1}{2m}p^{2}+\frac{m\omega^{2}}{2}\left(x+\frac{g_{k}}{m\omega^{2}}\right)^{2}\right.\\
& &\hspace{3em}+i\hbar\left(m\omega^{2}\left(x+\frac{g_{k}}{m\omega^{2}}\right)\frac{\partial}{\partial p}-\frac{1}{m}p\frac{\partial}{\partial x}\right)\nonumber\\
& & \hspace{12em}\left.+\epsilon_{k}-\frac{g_{k}^{2}}{2m\omega^{2}}\right]\Psi^{(h)}_{k} \nonumber
\end{eqnarray}
for the hybrid wave function $\Psi^{(h)}_k(q,p,t)$, defined over phase space.
%\block{(b) \emph{manipulations:}} 
Similar to the quantum case, for each $k$ we get a solution of the equation
$i\hbar \dot\Phi^{(c)}_k=\hat {\mathcal{O}}_{H_0} \Phi^{(c)}_k$ for the
classical harmonic oscillator in terms of the classical van Hove-wave function $\Phi^{(c)}_k(q,p,t)$ 
that for each $k$ has an offset energy and is displaced in position space:
\begin{eqnarray}
&&\Psi^{(h)}_k(q,p,t)=\\
&&\qquad\mathrm{e}^{-\frac{i}{\hbar}\left(\epsilon_k-\frac{g_k^2}{2m\omega^2}\right)t}\, \Phi^{(c)}_k\left(q+\frac{g_k}{m \omega^2},p,t\right)\nonumber\,.
\end{eqnarray}

To resemble the coherent state, used in the quantum case, we chose a Gaussian state $\Phi^{(c)}_\Sigma(q,p,t;q_0,p_0)$ of width $\Sigma$ centered around the initial position $q_0$ and a corresponding width $m\omega\Sigma$ around the momentum $p_0$.
%\block{(d) \emph{initial condition:}}  
As in the quantum case, at $t=0$ the state of the oscillator is measured, regardless of spin states, requiring the initial condition to be shifted separately for each of the spin wave function components: 
\begin{eqnarray}   &&\Phi^{(c)}_k\left(q+\frac{g_k}{m \omega^2},p,t\right)=\\
&&\qquad\qquad\Phi^{(c)}_\Sigma\left(q+\frac{g_k}{m \omega^2},p,t;q_0+\frac{g_k}{m \omega^2},p_0\right).\nonumber
\end{eqnarray}
For details about the van Hove wave function $\Phi^{(c)}_\Sigma(q,p,t;q_0,p_0)$ describing the evolution of the classical harmonic oscillator we refer to the general remarks at the beginning and to its explicit form given in the End Matter of this Letter, as well as to our previous papers \cite{Bermudez2024,Reginatto2025} for further reading.
 
The marginalization of the classical harmonic oscillator is performed by the phase-space integral
\begin{equation}  \rho^{(h)}_{ij}(t)=\int\mathrm{d}q\mathrm{d}p\,\Psi^{(h)\ast}_i(q,p,t)\Psi^{(h)}_j(q,p,t)\,.\label{eqn:h_margnial2}
\end{equation}
Applying the 
condition $\tau\simeq \tau'+t$ in the wave functions, Eq.~(\ref{eqn:h_margnial2}) simplifies to a Gaussian integral
over $q$ and $p$ that can be solved straightforwardly using standard methods. The resulting density matrix for the spin states is of the same form as in the quantum case (\ref{eqn:dm_01}) - (\ref{eqn:dm_03}), only deviating in the explicit expressions for the three functions
\begin{eqnarray}
 R^{(h)}(t)&=&\frac{g^2}{m^2  \omega ^4\Sigma ^2}(1-\cos ( \omega t))+\frac{1}{8}\frac{g^2 \Sigma ^2}{\omega ^2 \hbar ^2}\label{eqn:hdm_Function_01}
 \\
 S^{(h)}(t)&=&\frac{1}{2}\frac{g ^2 }{ m \omega ^3 \hbar }(\sin (\omega t )/2-\omega t)
 \\
 U^{(h)}(t)&=& 
\frac{g q_0}{\omega  \hbar}   \sin (\omega t) - \frac{gp_0}{m \omega ^2 \hbar}  \cos ( \omega t)+2\epsilon t/\hbar\,
\end{eqnarray}
now with the superscript `$(h)$' for \emph{hybrid} to denote that the two spins' (qubits') interaction now is mediated by the classical harmonic oscillator. 

To compare our hybrid result directly to the quantum case, in what follows, we set the width of the classical distribution to $\Sigma=\sqrt{\hbar/m\omega}$. In this case, the classical distribution becomes identical to the Wigner distribution of a coherent quantum state \cite{Case2008}. Moreover, it is natural to assume a finite width for the classical sector of a hybrid system, as the analysis of DeWitt \cite{DeWitt1962} points to a transfer of uncertainty that leads to limitations on the sharpness of the corresponding classical distribution \cite{Albers2008}.

\sectionPRL{Measures of entanglement} 

The derivation of similar density matrices for HvH and the fully quantum case can be seen as the major result of this work. From their explicit forms, which resemble each other, similar properties, such as entanglement, can be concluded in both cases. To further illustrate this, here we restrict ourselves to a brief discussion of two exemplary measures of entanglement, the purity and the concurrence, well knowing that many possible measures can be obtained from the density matrix (\ref{eqn:dm_01}), cf. Ref. \cite{4xHorodecki2009,Oh2006,Bengtsson2006}.

\begin{figure}[t!]
    \centering  \vspace{-0.5em} 
    \includegraphics[scale = 0.50]{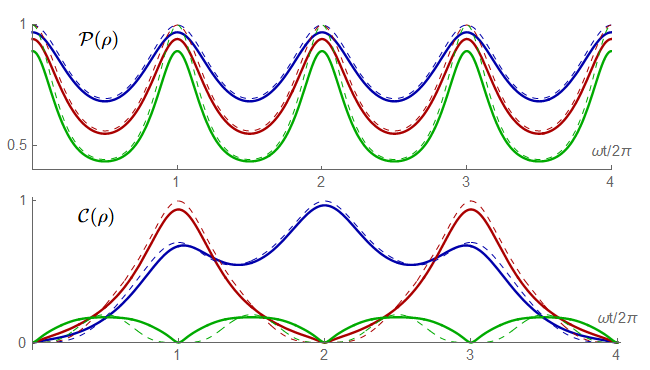}
    \caption{Comparison of the purity $\mathcal{P}(\rho)$ and the concurrence $\mathcal{C}(\rho)$ in the quantum case (dashed) and hybrid case (solid) with $\Sigma=\sqrt{\hbar/m\omega}$, respectively.
    Both measures are plotted for the interaction strength of $g^2/m \omega^3 \hbar=1$ (green), $g^2/m \omega^3 \hbar=1/2$ (red), and $g^2/m \omega^3 \hbar=1/4$ (blue). We find that the measures of entanglement for both quantum and hybrid systems closely resemble each other, especially when the interactions are weak ($g^2\ll m \omega^3\hbar$).  }
\label{fig:purity_and_concurrence}\vspace{-1em}
\end{figure}

%\skip{(b) \emph{purity definition and plot:}}
The \emph{purity} of the two spin system, plotted in Fig.~\ref{fig:purity_and_concurrence} is given by
\begin{equation}
\mathcal{P}(\rho)=\mathrm{tr}(\rho^2)=\frac{1}{8}\left(3+4\mathrm{e}^{-R}+\mathrm{e}^{-4R}\right)
\end{equation}
and is solely determined by the function $R$, that is Eq.~(\ref{eqn:qdm_Function_01}) for the quantum, and Eq.~(\ref{eqn:hdm_Function_01}) with $\Sigma=\sqrt{\hbar/m\omega}$ or the hybrid case.
The purity gives us information about correlations and to what extent the system is in a pure state, as indicated by $\mathcal{P}(\rho)=1$, coinciding with $R=0$. While that value is periodically reached by the quantum system, the hybrid system never is in a pure state, but approaches this case closely, if the interactions $g^2\ll m \omega^3\hbar$ are small.

We further calculate the \emph{concurrence} to investigate how entanglement builds up in the system. 
It is constructed from the density matrix $\rho$ and its spin-flipped version $\tilde\rho=S\rho^\ast S$, which can be obtained either by conjugation with $S=\sigma_2\otimes\sigma_2$  \cite{Wootters2001} or by flipping the Bloch vector $\vec{A}\to -\vec{A}$ in the Bloch-Fano decomposition (\ref{eqn:dm_01}), cf. Ref.~\cite{Bengtsson2006}. 
The concurrence then is obtained by 
\begin{equation}
\mathcal{C}(\rho)=\mathrm{max}(0,\lambda_1-\lambda_2-\lambda_3-\lambda_4)\,,\end{equation}
where $\lambda_i$ are the square roots of the eigenvalues of $\rho\tilde \rho$ in descending order.
In Fig.~\ref{fig:purity_and_concurrence}  we see that for small interaction strength $g^2\ll m \omega^3\hbar$ entanglement in the hybrid system is built up over several oscillation periods and closely follows the corresponding curve for the fully quantum system \footnote{See also Ref.~\cite{Oh2006}, where a comparable behavior of the concurrence for a similar quantum system is reported. Also notice the other measures of entanglement discussed therein.}.
In general, the concurrence gives a lower bound to \emph{entanglement of formation} \cite{Wootters2001,4xHorodecki2009,Bengtsson2006}. This also demonstrates \emph{distillable entanglement} in the hybrid system, since all entanglement in a two-spin (two-qubit) system is distillable \cite{4xHorodecki2009}.

\sectionPRL{Summary and Discussion}
We considered the problem of entanglement by a classical mediator by examining the example of two quantum spins coupled by a classical oscillator in the framework of HvH theory. In this hybrid theory, formulated in Hilbert space, the classical observables are represented by van Hove operators, which are well known in the context of geometric quantization \cite{Kirillov2001,G2000}. \changed{In our case, however, they are used to reformulate classical physics in Hilbert space. 
We have shown the equivalence between this formulation and the usual phase space description of classical physics in Ref.~\cite{Bermudez2024}.}

The theory has novel features which are lacking in the more familiar Koopman-von Neumann-Sudarshan approach \cite{K1931,vN1930,Sudarshan1976}. The commutators of the van Hove operators are isomorphic to the Poisson algebra of functions in phase space, which means that the algebraic structure of the classical observables of Hamiltonian mechanics is preserved; this, however, does not lead to an uncertainty relation for classical particles due to the absence of a product algebra for van Hove operators. \changed{Furthermore, no entanglement can be generated between classical particles in the classical sector of a hybrid system, since the classical subsystem is always described by a positive phase space distribution once the quantum sector is traced out. As shown by Fine \cite{Fine1982}, this is equivalent to the statement that the Bell inequalities hold for the classical sector (see End Matter for more details).} The well known problem of the phase of the classical wavefunction \cite{Sudarshan1976,M2002} is addressed by introducing conditions, Eq. (\ref{constr}),
which determine its functional form, Eq. (\ref{sigmaConstr}) \footnote{For a different resolution of the problem of the classical phase which does not rely on fixing the phase, see Ref. \cite{GAYBALMAZ2022133450}.}. In consequence, the van Hove operators are both observables and generators, unlike in Koopman-von Neumann theory. 
While van Hove operators have been used previously for hybrid systems, particularly in the Koopman-van Hove approach \cite{Bondar_2019,gaybalmaz:hal-03042578,GAYBALMAZ2022133450}, the HvH theory provides a new, consistent approach where important technical issues, e.g., equations of motion, handling of the phase of the wave function, and the definition of densities, are treated differently.

Applying the HvH formalism to the problem of two quantum spins, i.e., qubits, interacting via a classical harmonic oscillator, we find entanglement between the spins, similar to that of a purely quantum system. This is quantified by calculating the concurrence. The time evolution of the purity is also similar in both the hybrid and purely quantum cases. 

Our results differ from those obtained for similar systems but with other hybrid approaches, e.g. those of Elze \cite{Fratino2014} and Koopman-von Neumann \cite{GonzalezConde2023}, showing that the occurrence of entanglement depends on how the hybrid system is modeled \cite{HallReginatto2018}. Due to their close relationship to HvH theory \cite{Bermudez2024}, similar results are expected if the system is studied within the approach of ensembles on phase space \cite{Bermudez2024} and the approach of ensembles on configuration space \cite{HR2016}.

Our results show that the no-go theorems which deny the possibility of entanglement via a classical mediator, do not apply in general, as HvH theory provides a counterexample. This points to the fact that the assumptions of these theorems are too restrictive and do not cover all possible classical-quantum hybrid theories. 
\changed{For example, the no-go theorem of Marletto and Vedral \cite{MarlettoVedral2017} considers hybrid models where the classical observables are represented by commuting operators, but this assumption does not apply to our model.
A similar critique applies to the no-go theorem of Bose et al \cite{Bose2017} as it assumes that 
the mediator acts a classical communication channel within the framework of 
\emph{local operations and classical communication} (LOCC).
However, this kind of classical communication between the quantum systems can always be represented by 
communing operators in a sufficiently large Hilbert space that includes the two quantum systems and the mediator \cite{HallReginatto2018}.
Moreover, no-go theorems, which are based on the assumption that the hybrid theory is a \emph{generalized probabilistic theory} (GPT) do not apply to our model, just as they do not apply to the hybrid model of ensembles in configuration space \cite{HR2016,Galley2022}, because neither of them fits within the GPT framework. See also the End Matter for further discussion.}
A systematic and more detailed analysis of these and other no-go theorems \cite{PhysRevA.63.022101,PhysRevA.85.022127} and their applicability to HvH theory is the subject of
Ref.~\cite{Reginatto2026}.

We explicitly demonstrated that entanglement can be mediated via the classical sector of a hybrid theory.
Thus, also whether classical gravity, acting as a classical mediator, can lead to the entanglement of quantum systems remains an open question.
Our investigation, however, implies that the classical nature of gravity would still not be ruled out conclusively if entanglement of quantum systems by gravity can be demonstrated in experiments.

\sectionPRL{Acknowledgments}
S.U. acknowledges the support by the Deutsche Forschungsgemeinschaft (DFG, German Research Foundation) under Germany’s Excellence Strategy – EXC 2123/2 QuantumFrontiers – 390837967.
%the Deutsche Forschungsgemeinschaft (DFG, German Research Foundation) under Germany’s Excellence Strategy-EXC-2123 “QuantumFrontiers” -- Grant No. 390837967.
\bibliography{main}
\clearpage
\section{End Matter}
Here, we shed more light on classical wave functions, used in hybrid van Hove theory.
As in the main text, we restrict our discussion to one-dimensional single-particle systems and discuss the classical harmonic oscillator as an example case. The generalization to more dimensions and particles is straightforward. 

The classical wavefunction $\phi$ can be written in the form $\phi=\sqrt\varrho e^{i\sigma/\hbar}$, where the classical probability density $\varrho$, and the phase $\sigma$ satisfy Eqs.~(\ref{LE})-(\ref{dsigmadt}). 

\sectionPRL{The classical probability densities as a decomposition of trajectories}
We consider a one-dimensional harmonic oscillator with Hamiltonian $H_0=\frac{1}{2m}p^2+\frac{1}{2}m\omega^2q^2$. Let us further introduce the phase space functions
\begin{eqnarray}
\tilde Q(q',p',t) &=& q'\cos(\omega t)+\frac{p'}{m\omega}\sin(\omega t), \label{eqn:Qtilde}\\
\tilde P(q',p',t) &=& p'\cos(\omega t)-q' m \omega\sin(\omega t) \label{eqn:Ptilde}
\end{eqnarray}
that denote the classical trajectory that passes through the point $(q',p')$ at time $t=0$.
It is straightforward to check that the classical probability density $\rho$, that solves the Liouville equation  (\ref{LE}), can always be represented as a mixture of trajectories,
\begin{equation}\label{rhoTrajectories}
	\varrho(q,p,t) =  \sum_{q',p'} \, w_{q',p'}\delta(q-\tilde Q)\delta(p-\tilde P),
\end{equation} 
with $w_{q',p'} \ge 0$ and $\sum_{q',p'} \, w_{q',p'}=1$. In the case of a continuous distribution, the sum is replaced by an integral, $\sum_{q',p'} w_{q',p'} \rightarrow \int d q' d p' w(q',p')$. In particular, in the case of a single oscillator trajectory, we have
\begin{equation}\label{rhoOneTrajectory}
	\varrho'(q,p,t) = \delta(q-\tilde Q(q',p',t))\delta(p-\tilde P(q',p',t)) 
\end{equation} 
describing a particle that passes through the point $(q',p')$ at $t=0$. In what follows, we will also use the prime notation to denote quantities, which belong to a single, particular trajectory leading trough a point $(q',p')$ in phase space.
\sectionPRL{Requirements for the phase of the classical wave function}
We now consider Eq.~(\ref{dsigmadt}), which governs the evolution of the phase $\sigma$. 
Note that, starting from a solution $\sigma$ of Eq.~(\ref{dsigmadt}), there is a freedom to
write down new solutions $\sigma_F=\sigma + K(H_{0},\tau-t)$, where $K$ is an arbitrary function of the Hamiltonian $H_0$ and the phase space function $\tau(q,p)-t$, that is determined by the relation $\{\tau, H_0\}=1=d \tau/dt$. 

In HvH theory, $\sigma$ is fixed by the requirement that the expression for the expectation value of a van Hove operator corresponds with the phase space average over the corresponding phase space function, as stated by Eq.~(\ref{ExpValOperator}). 
Calculating the expectation value
\begin{equation}\label{average}
    \langle \hat{\mathcal{O}}_{F} \rangle =
    \int d q d p \, \mathcal{\varrho}\left[F+\left(\frac{\partial \sigma}{\partial q} - p \right) \frac{\partial F}{\partial p} - \frac{\partial \sigma}{\partial p} \frac{\partial F}{\partial q} \right],
\end{equation}
we find that this requirement holds by imposing the constraints of Eq.~(\ref{constr}). 
As we have shown in Appendix A of Ref.~\cite{Reginatto2025}, the solution of Eq.~(\ref{ExpValOperator}) that satisfies this constraints has the functional form of Eq.~(\ref{sigmaConstr}), i.e. $\sigma =  \eta + H_0[\tau- \tau'-t]$, where $\tau'$ is an initial value and $\eta(q,p)$ satisfies the relation $\{\eta,H_0\}=L=d \eta/dt$, which is consistent with the identification of $\eta$ as the classical action along a trajectory. Recalling that $\tau$ integrates to $\tau\simeq \tau'+t$, we also find that the phase $\sigma\simeq \eta$ coincides with the classical action in that case.

For the example of the one-dimensional harmonic oscillator we calculate 
\begin{equation}
\eta= \frac{qp}{2},\qquad \tau=\frac{1}{\omega}\tan^{-1}\left(\frac{m\omega q}{p}\right), \label{eqn:etatau}
\end{equation}
and can explicitly check that the constraints 
\begin{eqnarray}
	\frac{\partial {\sigma}}{\partial q} &\simeq &  \frac{p}{2} + \left.\frac{\partial}{\partial q}\left[H_0 (\tau-\tau'-t)\right]\right|_{\tau\simeq \tau' +t}\\ & = & \frac{p}{2} + H_0\frac{p}{2H_0} = p,\nonumber \\
    \frac{\partial {\sigma}}{\partial p} &\simeq & \frac{q}{2} + \left.\frac{\partial}{\partial p}\left[H_0 (\tau-\tau'-t)\right]\right|_{\tau\simeq \tau' +t} \\
    & =&  \frac{q}{2} - H_0\frac{q}{2H_0} = 0\nonumber
\end{eqnarray}are satisfied as required.

\sectionPRL{On classical wavefunctions for single trajectories}
To a single oscillator trajectory through $(q',p')$ we can associate a wavefunction
\begin{equation}\label{wftr1}  \phi'=\sqrt{\varrho'}\,e^{i\sigma'/\hbar}
\end{equation}
with the density $\varrho'$ from Eq.~(\ref{rhoOneTrajectory}) and $\sigma'=\eta + H[\tau-\tau'-t]$, where $\tau'=\tau(q',p')$.

Here we make the observation that the amplitude of the wave function in Eq.~(\ref{wftr1}) is formally expressed in terms of the square root of the delta functions from Eq.~(\ref{rhoOneTrajectory}). This does not present a conceptional difficulty, since the expectation values given by Eq. (\ref{ExpValOperator}) are always well defined, as one can see by either doing the transformation to Madelung variables or by reformulating classical mechanics using the approach of ensembles on phase space, which is equivalent to the wave function representation discussed here \cite{Bermudez2024}.  
Also, to prevent the square root of delta functions, $\sqrt{\rho}'$ can be defined properly using a limiting procedure.

Furthermore, since we are considering deterministic phase space trajectories, these trajectories cannot cross, otherwise the motion would not be deterministic at the point in which they intersect. This means that 
for two different initial conditions $(q',p') \neq (q'',p'')$ the corresponding densities $\varrho'$ and $\varrho''$ are defined on two disjunct regions of phase space, if they belong to different trajectories of the same classical system. Moreover, particles with different initial conditions, traveling on the same trajectory can not meet, due to the fixed time span $\tau''-\tau'$ between them. Consequently, the wave functions
\begin{equation}\label{nooverlap}
\int dqdp\, \phi'^\ast(q,p,t)\,\phi''(q,p,t)=0   
\end{equation}
are \textit{orthonormal}. 
Finally, since $\tau\simeq\tau'+t$ when $\tau$ is evaluated on a trajectory, the \textit{numerical value} of the wavefunction simplifies to
\begin{equation}\label{valueClassicalWaveFunctionTrajectoriesW} \phi' \simeq  \sqrt{\varrho'}\,e^{i\eta/\hbar},
\end{equation}
with a phase that does not depend any longer on the initial conditions $(q',p')$. Moreover, the phase is equal to the action $\eta$
for all trajectories of the system.
This allows us to introduce a representation of the state in terms of a wavefunction associated with any density $\varrho$ that solves the Liouville equation given by
\begin{equation}\label{wf1w} \phi \simeq \sqrt{\varrho\,}\,e^{i\eta/\hbar}.
\end{equation}
One should keep in mind that 
the numerical values
(\ref{valueClassicalWaveFunctionTrajectoriesW})-(\ref{wf1w}) can only be applied \textit{after} all manipulations involving van Hove operators acting on states have been carried out, 
as for instance, it was done in  Eq.~(\ref{eqn:h_margnial2}) before integration.
Otherwise expressions 
which rely on the full functional form of the wave function,
like Eq.~(\ref{average}), 
would
not evaluate correctly. This is analogous to the handling of weak equalities in Dirac's theory of constraints \cite{Dirac:2001:LQM}.

\sectionPRL{Coherent state for the quantum oscillator and its classical analogue}

The coherent state $\phi_c(q,t;q_0,p_0)$ for the quantum oscillator \cite{CT1977} is given by
\begin{eqnarray}\label{cseq}
\phi_c &=& \left(\frac{m\omega}{\pi \hbar}\right)^{1/4}\mathrm{exp}\Big\{-\frac{m\omega}{2\hbar}[q-Q(t)]^2 \nonumber\\
&~& \qquad \qquad \qquad \qquad +\frac{i}{\hbar} P(t)q-i \theta(t)\Big\}\quad
\end{eqnarray}
with 
the time-dependent functions
\begin{eqnarray}
    Q(t)&=&q_0 \cos(\omega t)+ \frac{p_0}{m\omega} \sin(\omega t),\\
    P(t)&=&p_0 \cos(\omega t)- m\omega q_0 \sin(\omega t),\\
    \theta(t)&=&\frac{\omega t}{2} +\frac{1}{2 \hbar} Q(t)P(t).
\end{eqnarray}

The 
van Hove wave function $\Phi^{(c)}_\Sigma(q,p,t;q_0,p_0)$ for a Gaussian 
state of a
classical oscillator, 
which must satisfy $i\hbar \dot\Phi^{(c)}_\Sigma=\hat {\mathcal{O}}_{H_0} \Phi^{(c)}_\Sigma$, where 
$H_0=\frac{1}{2m}p^2+\frac{1}{2}m\omega^2q^2$
is the classical harmonic oscillator Hamiltonian
, takes the form
\begin{eqnarray}\label{vhacseq}
\Phi^{(c)}_\Sigma&=&\frac{1}{\Sigma\sqrt{\pi m \omega}}\mathrm{exp}\Big\{-\frac{1}{2\Sigma^2}[q_0-\tilde{Q}(q,p,t)]^2\nonumber\\
&~& \left.-\frac{1}{2(m\omega\Sigma)^2}[p_0-\tilde{P}(q,p,t)]^2+i \eta(q,p)/\hbar\right.\nonumber\\
&~& - i H_0(q,p)[t+\tau_0-\tau(q,p)]/\hbar\Big\}\,,
\end{eqnarray}
where 
the functions $\tilde Q(q,p,t)$ and $\tilde P(q,p,t)$ are given by Eqs.~(\ref{eqn:Qtilde}) and (\ref{eqn:Ptilde}), and $\eta(q,p)$ and $\tau(q,p)$ are given by Eq.~(\ref{eqn:etatau}).
The Gaussian state of width $\Sigma$ is initially centered around the position $q_0$ and a corresponding width $m\omega\Sigma$ around the momentum $p_0$.
For the particular choice $\Sigma=\sqrt{\hbar/m\omega}$
the classical probability distribution 
$\rho^{(c)}_\Sigma=|\Phi^{(c)}_\Sigma|^2$ equals the Wigner function for a comparable quantum coherent state \cite{Case2008}.
\changed{\sectionPRL{Absence of non-classical correlations for classical subsystems in HvH theory} In the main text of the Letter we show that in HvH theory two quantum systems can become entangled by interaction via a classical mediator. However, it is natural to ask, what prevents potential non-classical correlations (i.e., violating Bell inequalities) in the opposite case of two classical systems interacting via a quantum mediator:}

\changed{
As an example, we assume two classical particles coupled to a quantum spin with the hybrid Hamiltonian
\[
\hat{\mathcal{H}}=\hat{\mathcal{O}}_{H_1(q,p)}+\hat{\mathcal{O}}_{H_2(q',p')}+k(\hat{\mathcal{O}}_{F_1(q,p)}+\hat{\mathcal{O}}_{F_2(q',p')})\,\sigma_3\,,
\]
where we keep $H_1$, $H_2$, $F_1$ and $F_2$ as general as possible. As in the main text of the Letter, we obtain a hybrid wave function 
$\Psi=(\phi^{(c)}_{\uparrow}\,,\,\phi^{(c)}_{\downarrow}\,)/\sqrt{2}$
with each of the components related to the spin states, which acts like a classical system with the Hamiltonian
\[
H_{\uparrow/\downarrow}=H_1+H_2\pm k (F_1+F_2)\,,
\]
respectively. 
The hybrid density matrix $\rho_{ij}=\Psi^\ast_i\Psi_j$, after marginalization of the quantum sector, yields
$\rho_{cs}=\mathrm{tr}\rho=\left(\rho_{\uparrow}+\rho_{\downarrow}\right)/2\,,
$
where the classical density matrices $\rho_{i}=\phi_i^{(c)\ast}\phi^{(c)}_i$ evolve according to
\[
\{H_\uparrow,\rho_\uparrow\}=\dot\rho_{\uparrow}\quad,\quad \{H_{\downarrow},\rho_\downarrow\}=\dot\rho_{\downarrow}\,.
\]
Hence, for the hybrid model, the evolution of the classical subsystem is not guided by a single Hamiltonian acting on a single phase space density, as it would be the case for a purely classical system.
Instead we find a mixture of two classical systems, associated with the states ($\uparrow/\downarrow$) of the quantum spin. 
This also indicates that HvH theory does not fit within 
the framework of GPTs, where one would expect the evolution of the classical subsystem to follow \sout{a} the usual reduced description given in terms of a single Liouville equation.}

%which require the evolution of statistics to be determined by a 
%This, \textcolor{blue}{incidentally, suggests} that HvH does not fit within the framework of \textcolor{orange}{generalized probability theories (GPTs), }\sout{: if this were the case, one would expect the evolution of the classical subsystem to follow a reduced description given in terms of a single Liouville equation.}
%\textcolor{orange}{where it is assumed that the classical subsystem follows a single Liouville equation.}}

\changed{To make clear, that no non-classical correlations violating Bell inequalities are possible between the particles with coordinates $(q,p)$ and $(q',p')$, we point to the fact that both $\rho_{\uparrow/\downarrow}(q,p;q',p')$ are by definition ($\rho_i=|\phi_i^{(c)}|^2$) always positive phase space probability densities. The same holds also for $\rho_{cs}$, as their sum.
In fact, $\rho_{cs}>0$ is generally the case in hybrid van Hove theory. The crucial point is that a system with a positive phase space probability density is incapable of violating Bell's inequality, as first demonstrated by Fine in his characterization of the Bell inequalities in terms of joint probability distributions \cite{Fine1982,Auberson2002}. In consequence, no quantum entanglement can occur between the classical particles in the Hybrid system, described by HvH theory.}

\end{document}